

\magnification=\magstep1
\openup 2\jot
\def\bo{ { \sqcup\llap{ $\sqcap$} } }
\overfullrule=0pt       



\def\hrr{{h^{rr}}}
\def\htr{{h^{tr}}}
\def\Lam2{\Lambda^2}

\def\real{I\negthinspace R}

\def\half{\textstyle{1\over2}}
\def\quarter{\textstyle{1\over4}}

\def\r3{I\negthinspace R$^3$}


\hyphenation{
mini-su-per-space
pre-factor
tei-tel-boim
es-po-si-to
haw-king
Bala-chand-ran
}

\def\pois{1}
\def\gibmae{2}
\def\grehar{3}
\def\papone{4}
\def\paptwo{5}
\def\hs{6}

\hbox{ }
\rightline {DAMTP/R-94/40}
\rightline {LA-UR-94-3323}
\vskip 1truecm

\centerline{\bf EVIDENCE FOR STABILITY OF EXTREMAL BLACK p-BRANES}
\vskip 1truecm

\centerline{Ruth Gregory}
\vskip 2mm

\centerline{ \it D.A.M.T.P. University of Cambridge}
\centerline{ \it Silver Street, Cambridge, CB3 9EW, U.K.}

\vskip 4mm
\centerline{and}
\vskip 4mm

\centerline{ Raymond Laflamme }
\vskip 2mm
\centerline{\it Newton Institute for the Mathematical Sciences, University of Cambridge}
\centerline{\it 20 Clarkson Road, Cambridge, U.K. CB3 0EH}
\vskip 1mm
\centerline{\it and}
\vskip 1mm
\centerline{ \it Theoretical Astrophysics, T-6, MSB288,
Los Alamos National Laboratory}
\centerline{ \it  Los Alamos,  NM 87545,USA}

\vskip 4mm
\centerline{Abstract}

\noindent We investigate the stability of the extremal 
black p-brane which contains
a n-form and a dilaton.  We show that the instability due to the s-mode,
which was present in the uncharged and non-extremal p-brane, disappears
in the extreme case.  This is shown to be consistent with an entropy argument
which shows that the zero entropy of the extremal black hole is approached 
more rapidly than the zero entropy of the black p-brane, which would mean
an instability would violate the  second law of thermodynamics.

\openup 1\jot


\vfill\eject
\footline={\hss\tenrm\folio\hss}

\noindent{\bf 1. Introductory remarks.}

\vskip 2mm

In four dimensions, black holes are characterised by a small set of parameters: 
charge, mass and angular momentum. The simplest solution, Schwarzschild,
representing an uncharged non-rotating black hole has long been known to be
classically stable, and in Einstein gravity, the stability of the exterior
spacetime of the charged and rotating counterparts (Reissner-Nordstrom,
Kerr-Newman) has also been demonstrated. In this latter case, there is an
instability of the inner Cauchy horizon, however, this is hidden from us by the
event horizon, and is therefore presumably not of physical interest to the
exterior observer. Therefore, once a black hole is formed, it remains, forever
shielding the details of its constituents from the outside observer, who may
only measure the charge, mass, and angular momentum of that matter which formed
the black hole. 

Quantum mechanically, we have a very different picture. Black holes become
black bodies, semi-classically radiating a purely thermal spectrum. In theory, a
black hole could radiate away completely, possibly losing forever all the
detailed information of its constituents. Such a picture, in which quantum
mechanical unitarity is lost, is potentially unsettling, and certainly begs the
question as to whether there is not some modification of semi-classical gravity
in which information retrieval is in some way possible. An examination of black
holes in low energy string theory is a step in this direction. Here, the
Einstein action is modified by the addition of a dilaton with a non-trivial
coupling to `electromagnetism'
$$
S = \int d^4x \sqrt{-g} [R - 2(\nabla\phi)^2 - e^{-2\phi} F^2] \ .
\eqno (1.1)
$$
Such `dilatonic' black holes have several attractive features. In
Einstein gravity, charged black holes  have an instability of the
inner Cauchy horizon due to matter perturbations in the exterior
spacetime$^\pois$, however, there is no static charged black hole solution in
Einstein gravity with only one horizon and a spacelike singularity.
On the other hand, in low energy string gravity the dilaton greatly changes 
the causal structure of charged black holes making them like Schwarzschild,
having  one event horizon and a spacelike singularity$^\gibmae$. 
This structure is generic, even if the dilaton has a mass$^\grehar$,
 as is generally believed from,
although not necessarily implied by, the principle of equivalence.
Additionally, extremal magnetically charged dilaton black holes have an
infinite throat, which, unlike the spatially infinite throat of extremal
Reissner-Nordstr\"om, is a true internal infinity as seen in the so-called
string metric
$$
g_{_{\rm string}} = e^{-2\phi} g_{_{\rm Einstein}}
\eqno (1.2)
$$
Such an internal infinity allows an effective dimensional reduction down to
two-dimensions in the throat and has been a topic of much debate as a
simplified model for black hole accretion/radiation.

Since superstring theory is most naturally, though not necessarily, associated
with a ten-dimensional background, ideally one should examine string gravity in
higher dimensions. In  four dimensions, an event horizon must be topologically
spherical, but in higher dimensions this need not be the case, for example, in
ten dimensions we can have $S^2\times$ \real$^6$, $S^3 \times$ \real$^5$
{\it etc.}, topologies for the horizon. In previous
work$^{\papone,\paptwo}$,  we have pointed out
that such black p-branes are unstable for a large range of (magnetic) charge:mass
ratios, however, we were unable to demonstrate an instability for the extremal
black p-branes. This was largely because our numerical analysis was tailored
towards a general charge:mass ratio, and if ``Q=M'', a degeneracy occurs in the
perturbation equations reducing the dimensionality of the numerical problem
from 5 to 3, this degeneracy causing numerical instability in the solution of
the equations. In this paper, we seek to close this loophole in our argument,
tailoring our analysis specifically to extremal black p-branes, providing a
convincing argument for the stability of such objects.

The layout of  the paper is as follows. We first review some heuristic
arguments, particularly a thermodynamic entropy argument, as to the stability
(or otherwise) of black p-branes. We then set up the formalism involved in
solving the problem and argue that the only possible instability is an s-mode.
We then examine this s-mode numerically and show there is no solution to the
perturbation equations corresponding to an unstable mode. Finally we summarise
our results and discuss their implications.

We would first like to examine several arguments in favour of stability for
extremal black p-branes. It is generally believed that, being supersymmetric
and satisfying an appropriate Bogomolnyi bound, extremal black p-branes should
be stable, although formulating an appropriate argument in low energy string
gravity would be necessary to prove this. One of the main problems is that the
mass appearing in the Bogomolnyi bound is generally the mass per unit p-area,
and this is defined by taking the ADM mass of the D-dimensional black hole.
Such a procedure relies crucially on the translational invariance of the metric
in the p orthogonal directions. Since our instability, when it exists,
involves a dependence on these directions, it technically sidesteps the bound.
In order to close this route one would have to define a generalised notion of
`asymptotically flat', re-derive the ADM mass and positivity theorems, and
finally the Bogomolnyi bound for `approximately p-planar' objects. 

 A second consideration is that certain extremal p-branes  are relevant in the
context of S-duality - a generalised strong-weak coupling duality which, loosely
speaking, swaps solitons and gauge particles, electric and magnetic charges. If
a supersymmetric brane were unstable, it would be rather embarrassing for the
dual particle! Although this is not a good reason for stability, there is
sufficient circumstantial evidence in favour of a limited form of S-duality that
it would be most remarkable if we were to discover that the supersymmetric
branes were unstable. 

Finally, our main reason for expecting stability
 is a thermodynamic one. One of the most convincing arguments that we
should expect an {\it instability} of the uncharged black p-brane was an
entropic inequality. We calculated the entropy of a length $L$ of black string
in five dimensions and compared it to that of a hyperspherical black hole. For
the string, the entropy was proportional to $M^2L$, for the black hole,
$M^{3/2}L^{3/2}$. Regarding M, the mass per unit length of the string as a
constant, this showed that the entropy of the hole is greater for $L\ge {\cal
O}(M)$, and hence that a series of black holes is preferred to the string.
However, such an argument did not include charge and the non-trivial dilaton
field that this entails. It is therefore worth re-investigating this argument
and generalising it to include charge. We will perform this analysis for general
$N$, the number of spacetime dimensions, and general $D=N-$p, so as to be able
to show the genericity of the argument. 

In order to calculate the entropy of the black p-brane, it is most convenient
to work in the ``Einstein'' frame, {\it i.e.}, to make a conformal
transformation so that the gravitational part of the action is in Einstein form
$$
S = \int \! d^N x \sqrt{-g} R + ....
\eqno (1.3)
$$
The appropriate conformal factor for $N$ dimensions is
$$
g_{_{\rm Einstein}} = e^{-4\phi/(N-2)} g_{_{\rm string}}
\eqno (1.4)
$$
where $\phi$ is the dilaton field. Applying this conformal factor to the metric
of a p-brane yields:
$$
\eqalign{
ds^2_{_{\rm Ein}} = &- {\bigl ( 1 - ({r_+\over r})^{D-3}\bigr ) \over
\bigl ( 1 - ({r_-\over r})^{D-3} \bigr )^{N-4\over N-2} } dt^2 
+ {dr^2 \over \bigl ( 1 - ({r_+\over r})^{D-3}\bigr ) 
\bigl ( 1 - ({r_-\over r})^{D-3} \bigr )^{N-4\over N-2} } \cr
&+ r^2 \bigl ( 1 - ({r_-\over r})^{D-3} \bigr )^{2\over N-2} d\Omega^2_{D-2}
+ \bigl ( 1 - ({r_-\over r})^{D-3} \bigr )^{2\over N-2} dx_i dx^i \cr
}
\eqno (1.5)
$$
>From this, we read off the entropy using $S=\quarter A$ as
$$
S = \quarter A_{D-2} r_+^{D-2} \left ( 1 - ({r_-\over r_+})^{D-3} \right
)^{D-2\over N-2}
\eqno (1.6)
$$
where
$$
A_n = {2 \pi^{n+1\over 2} \over \Gamma \left ( {n+1\over 2} \right ) }
\eqno (1.7)
$$
is the area of a unit n-sphere.

In order to compare the entropies of a p{\sevenrm=(N-D)}-brane with an
N-dimensional black hole, we need the entropy as a function of mass and charge.
To find the mass of a D-dimensional black hole, note that as $r\to\infty$
$$
|g_{00}| \sim 1 - {16\pi\over (D-2) A_{D-2}} {M_D\over r^{D-3}}
\eqno (1.8)
$$
Reading off the asymptotic form from (1.5) gives
$$
M_D = {(D-2)A_{D-2} \over 16\pi} \left [ r_+^{D-3} - {\textstyle{N-4\over N-2}}
r_-^{D-3} \right ]
\eqno (1.9)
$$
Additionally, $Q_D^2 = {D-3\over2}(r_+r_-)^{D-3}$, which gives
$$
\eqalign{
r_+^{D-3} &= {8 \pi M_D \over (D-2)A_{D-2}} + \sqrt{
{64 \pi^2 M_D^2 \over (D-2)^2 A_{D-2}^2} + {2Q_D^2 (N-4)\over(N-2)(D-3)} } \cr
r_-^{D-3} &= {-8 \pi M_D (N-2) \over (D-2) (N-4) A_{D-2}} + \sqrt{
{64 \pi^2 M_D^2 (N-2)^2 \over (D-2)^2 (N-4)^2 A_{D-2}^2} + {2Q_D^2
(N-2)\over(N-4)(D-3)} } \cr
}
\eqno (1.10)
$$
Thus we input the appropriate form of $r_\pm$ into $S_D$ to find the entropy
per unit area of the black p-brane, and setting $D=N$ gives the relevant
quantities for the N-dimensional black hole. We then compare the entropies by
taking a `cube' of p-brane, side-length $L$, and setting
$$
\eqalign{
M_N &= L^{N-D} M_D \cr
Q_N &= \sqrt{(N-3)\over (D-3)} {(D-2)\over (N-2)} {A_{D-2}\over A_{N-2}} L^{N-D}
Q_D = \sqrt{2(N-3)} {4\pi c\over A_{N-2}} L^{N-D} M_D  \cr 
}
\eqno (1.11)
$$
The rather unusual factors appearing in the charge formula are to ensure that
the extremal limit of the black p-brane corresponds to the extremal limit of
the black hole. If this were not the case, then the p-brane would either become
stable before reaching its extremal limit (a scenario we have already
eliminated), or would be unstable at all length scales at its extremal limit,
and in particular at small length scales close to the extremal limit; this too
is ruled out by our previous work which showed that the length scale of the 
instability was tending to infinity as the extremal limit was approached. We
therefore assume that these limits coincide, and introduce the constant $c$ as
representing a general charge:mass ratio, so that $c\to1$ as the extremal limit
is approached. 

We now set $L^{N-D}S_D = S_N$ to find the critical length, $L_*$,  at which it
becomes entropically favoured to form an N-dimensional black hole. Taking
$2\pi/L_*$ gives $\mu_*$, the critical frequency which, after some algebra, is
found to be:
$$
\eqalign{
\mu_* = &{\textstyle {2\pi \over 4^{1\over D-3}} }
\left ( {\textstyle {N-4\over 2} }\right )^{(D-2)(N-3)\over (N-2)(N-D)} 
\left ( {\textstyle {\Gamma \bigl ({N-1\over2}\bigr) \over N-2 } }\right
)^{1\over N-D} 
\left ( {\textstyle {\Gamma \bigl ({D-1\over2}\bigr) \over D-2 } } \right
)^{-(N-3)\over (D-3)(N-D)} 
\left ( {\textstyle {2(D-2)\over (N-2)(N-4)} }\right )^{N-3\over N-D} \cr
&\times \left ( {\scriptstyle 1 + \sqrt {1+c^2(N-2)(N-4)} } \right
)^{-(ND-2N-3D+7)\over(N-2)(D-3)} 
\left ( {\scriptstyle N-3 - \sqrt {1+c^2(N-2)(N-4)} }\right)^{N-3\over N-2} \cr 
}
\eqno (1.12)
$$
Clearly
$$
\eqalign{
\mu_* \propto & (N-3-\sqrt {1+c^2(N-2)(N-4)})^{N-3\over N-2} \cr
&= (N-3-\sqrt {1- c^2 +c^2(N-3)^2})^{N-3\over N-2} \cr
&\to 0 \hskip 2cm {\rm as} \ \ \ c\to 1. \cr
}
\eqno (1.13)
$$
Figure 1. shows the numerical and analytical value for $\mu_{max}$
for which the instability just disappears.  It is seen that in both
cases $\mu_{max}$ goes to zero in the limit of extreme charge.
Hence, as the extremal limit is approached, the scale at which it becomes
entropically favourable to form a black hole tends to infinity. Thus the entropy
argument, which was so convincing for uncharged black holes, actually
reinforces the idea that extremal black holes should be stable.
We now turn to setting up the formalism for the stability analysis. Since this
has been done in detail in [\paptwo], we will merely paraphrase the process here,
quoting only the main formulae.

\vskip 3mm

\noindent{\bf 2. The stability analysis.} 

\vskip 2mm

The low energy string action used by Horowitz and Strominger$^{\hs}$ is
$$
\int d^{10}x \sqrt{-g} e^{-2\phi}[ R + 4(\nabla \phi)^2 - {2\over(D-2)!}F^2]
\eqno (2.1)
$$
where $F$ is a $(D-2)$-form field strength, and $\phi$ the dilaton. This has an
extremal magnetically charged black p-brane solution:
$$
\eqalign{
ds^2 &= -dt^2 + {dr^2 \over ( 1 - ({r_+\over r})^{D-3})^2} + r^2 d\Omega^2_{D-2}
+ dx^i dx^j \delta_{ij}\cr
e^{-2\phi} &= ( 1 - ({r_+\over r})^{D-3}) \cr
F &= \sqrt{D-3\over 2} r_+^{D-3} \epsilon_{D-2} \cr
}
\eqno (2.2)
$$
where the index $i$ runs from $D+1$ to 10 and $\epsilon_{D-2}$ is the area form
of a unit (D-2)-sphere.  
In order to write the perturbation equations, we use the usual notation
$$
\delta g_{ab} = h_{ab}\ ,
\eqno (2.3)
$$
and a conventional gauge choice of transversality
$$
\nabla_a \bar{h}^{ab} = \nabla_a (h^{ab} - \half h g^{ab}) = 0 \ .
\eqno (2.4)
$$
To perform the stability analysis, we Fourier decompose the perturbations
in terms of the symmetries of the background spacetime. We perform our
analysis in Schwarzschild coordinates, and transform to the tortoise
coordinate at the future event horizon to check the regularity of the
perturbation equation (see [\paptwo] for a more detailed discussion of the issue of
boundary conditions and the generalised tortoise coordinate).   The Fourier
modes in the time and p-brane coordinates are of the form $e^{\Omega t + i
\mu_ix^i}$ for an instability. The spherical harmonic modes will depend on the
number of dimensions, $D$, that the black hole sits in, as well as the tensorial
nature of the perturbation we are analysing. However, we note that in the
uncharged black p-brane, the higher angular momentum mode equations have no
unstable solutions, and also that the s-wave instability shrinks in
parameter range as we increase the charge (see figure 1.). 
In other words, adding charge
has a stabilising influence. Thus we do not expect higher angular momentum
perturbations of charged black holes to exhibit instabilities. Therefore,
for the extremal black p-brane, we focus on the s-mode which caused an
instability in the lesser-charged branes. This mode takes the form
$$
\eqalignno{
\delta \phi &= e^{\Omega t + i \mu_ix^i} f(r) & (2.5a) \cr
\delta F &= 0 &(2.5b) \cr
h^{ab} &= e^{\Omega t + i \mu_ix^i} \left [
{\matrix{ H^{ij}(r) & H^{it}(r) & H^{ir}(r) & 0&0 & .... \cr
H^{tj}(r) & H^{tt}(r) & H^{tr}(r) & 0 & 0 &.... \cr
H^{rj}(r) & H^{rt}(r) & H^{rr}(r) & 0 & 0 &.... \cr
0 & 0 & 0 & K(r) &0 & 0.. \cr
0 & 0 & 0 & 0 & K(r)/\sin^2\theta & 0.. \cr
..&..&..&..&..&..\cr}} \right ] & (2.5c) \cr
}
$$
where we have used the non-trivial result from [\paptwo] that $\delta F=0$.

Using the gauge conditions, and (2.5b), the perturbation equations in
tensorial notation are
 $$
\eqalignno{
& \bo \delta \phi - h^{ab}\nabla_a \nabla_b \phi +2 h^{ab} \nabla_a\phi
\nabla_b \phi
- 4 \nabla_a\delta \phi\nabla^a  \phi \cr
& \;\;\;\;\;  -{h^{cd}\over (D-4)!}
F_{ca_2...a_{D-2}} F_d^{\;a_2...a_{D-2}} =0 & (2.6a) \cr
&\bo h_{ab} + 2 R_{cadb}h^{cd} - 2 R_{e(a} h_{b)}^e 
 - 4 \nabla_a\nabla_b \delta\phi - 2 \nabla_c\phi \nabla^c h_{ab}
+ 4 \nabla_c \phi \nabla_{(b} h_{a)}^c \cr
& - {4\over (D-4)!}  h^{cd}
F_{aca_3...a_{D-2}} F_{bd}^{\;\;a_3...a_{D-2}} =0 & (2.6b) \cr
}
$$

Inputting the mode decomposition (2.5) yields the full set of equations which
were written down in appendix B of [\paptwo]. We therefore set $r_+=r_-$ in these
equations, and without loss of generality we set D=4, since we demonstrated
that the existence of an instability was not dimensionally dependent.
In this case, the equations of [\paptwo] simplify neatly. 
An important feature is that
the equations for $f,K,h^{rr}$ become independent of all the other variables.
It is in this sense that the extremal limit is degenerate, the number of 
independent physical variables in the perturbation problem reducing from
five to three.
We can therefore look first at these functions to see if the instability
persists. We should also point out that in the extreme case the behaviour of
the modes near $r_+$ is different than in the non-extreme case, this is because
of the effect of the equality of $r_-$ and $r_+$.  We do the analysis
for $D=4$ in 5 dimensions but it can be generalised for higher dimensions
and $D$.
The equations for these modes are:
$$
\eqalignno{
-{(r-r_+)^2\over r^2} f''-{(2r+r_+)(r-r_+)\over r^3} f' 
-{(r-2r_+)r_+\over r^2}K 
&\cr
-{r_+(3r_+-2r)\over 2r^2(r-r_+)^2} \hrr + \Lambda^2 f      
& = 0\;\;\; & (2.7a) \cr
-{(r-r_+)^2\over 2r^2} K'' + {(-3r+2r_+)(r-r_+)\over r^3} K' 
+ {2(r-r_+)^2 \over r^5}f'
&\cr
-{(r-3r_+)\over r^4(r-r_+)}\hrr + \big ({\Lam2\over 2} 
+ {(3r_+^2 -2r^2)\over r^4} \big)K
& =0 & (2.7b)\cr
-{(r-r_+)^2\over 2r^2} \hrr'' + {2(r-r_+)^4 \over r^4} f'' 
- {(r-3r_+)(r-r_+)\over r^3}\hrr'  + {2(r-r_+)^3 r_+ \over r^5} f'
&\cr
-{2(r-2r_+)(r-r_+)^3\over r^4} K + 
\big( {\Lam2 \over 2} + {(2r^2-6rr_+ + r_+^2)\over r^4} \big) \hrr 
& =0 \;\;\; & (2.7c) \cr}
$$
Using (2.7b,c) we can obtain
$$\eqalign{
-{(r-r_+)^2\over 2r^2} \hrr'' - {(r-3r_+)(r-r_+)\over r^3}\hrr' -{4(r-r_+)^3\over r^4} f'
&\cr
-{2(r-2r_+)(r-r_+)^2 \over r^3} K + {2\Lam2 (r-r_+)^2\over r^2}f
+\big ({\Lam2\over 2} + {2(r^2 -2rr_+ -r_+^2)\over r^4}\big) \hrr & =0 \cr}
\eqno (2.8)
$$
where $\Lambda^2 = \Omega^2 + \mu^2$.
We can see that these equations depend only on $\mu$ through 
$\Lambda$.  If we would find an unstable mode for a given
value of $\Lambda$ this would imply one for $\mu=0$ and $\Omega$ equal
to $\Lambda$.  This would correspond to an instability for a
4-dimensional dilatonic black hole!  However as will be shown below, no
such instability was found.

For general $\Lambda$, we cannot directly determine if there exists a regular
unstable solution. We must therefore resort to numerical techniques. These
involve finding the regular solution space at infinity, which forms a three
dimensional subspace of the total solution space. We then integrate in,
adjusting the initial configuration to see if one matches to a regular solution
near the horizon. In practise, as described in [\paptwo], 
we will search for evidence of such a match, rather than the match itself.

As $r\rightarrow\infty$ we can expand the regular solution as
$$
\eqalign{
f\rightarrow&\  f_i e^{\Lambda r} \cr
K\rightarrow&\ K_i e^{\Lambda r} \cr
h^{rr}\rightarrow&\ H^{rr}_i e^{\Lambda r} \cr}
\eqno (2.9)
$$
where $f_i, K_i, H^{rr}_i$ are arbitrary constants.  In fact 
one of them can be set to 1 due to the linearity of the equation.

In the limit as $r\rightarrow r_+$ the behaviour of the field is rather different
than in the non-extreme case. We assume the form  
$$
\eqalign{
f &= \bar f (r-r_+)^\alpha \cr
K &= \bar K (r-r_+)^\beta \cr
\hrr &= \bar \hrr (r-r_+)^\gamma \cr}
\eqno (2.10)
$$
and find that

$$\eqalign{
f=& {A_+}(r-r_+)^{\alpha_1^+} + {A_-}(r-r_+)^{\alpha_1^-} + {B_+}(r-r_+)^{\alpha_2^+}  \cr
  & {B_-}(r-r_+)^{\alpha_2^-} + {C_+}(r-r_+)^{\alpha_3^+} + {C_-}(r-r_+)^{\alpha_3^-}  \cr
\hrr =& -{2{\alpha_1^+}^2\over (\alpha_1^+ -1)r_+^2} {A_+}(r-r_+)^{\alpha_1^+ +2} 
        -{2{\alpha_1^-}^2\over (\alpha_1^- -1)r_+^2} {A_-}(r-r_+)^{\alpha_1^- +2} 
        -{4\alpha_2^+\over r_+^2}{B_+}(r-r_+)^{\alpha_2^+ +2}  \cr
  & -{4\alpha_2^-\over r_+^2}{B_-}(r-r_+)^{\alpha_2^- +2} 
    -{2\alpha_3^+\over r_+^2}{C_+}(r-r_+)^{\alpha_3^+ +2} 
    -{2\alpha_3^-\over r_+^2}{C_-}(r-r_+)^{\alpha_3^- +2}  \cr
K=&   {(\alpha_1^+ -1)\over (\alpha_1^+ -2)}{A_+}(r-r_+)^{\alpha_1^+} 
    + {(\alpha_1^- -1)\over (\alpha_1^- -2)}{A_-}(r-r_+)^{\alpha_1^-}
    - {4\over r_+^3} {B_+}(r-r_+)^{\alpha_2^+ +1}  \cr
  & - {4\over r_+^3} {B_-}(r-r_+)^{\alpha_2^- +1} 
    - {2\over r_+^3} {C_+}(r-r_+)^{\alpha_3^+ +1} 
    - {2\over r_+^3}{C_-}(r-r_+)^{\alpha_3^- +1}  \cr} 
\eqno (2.11)
$$
for which
$$
\eqalignno{
\alpha_1^\pm=\alpha=\beta=\gamma-2 &= -{1\over 2} \pm {3\over 2}
\sqrt{1 + {4\Lam2r_+^2\over 9}} & (2.12a) \cr
\alpha_2^\pm=\alpha =\beta-1 = \gamma-2 &= \pm\Lambda r_+  & (2.12b) \cr
\alpha_3^\pm=\alpha=\beta-1=\gamma-2 &= -{1\over 2} \pm {1\over 2}
\sqrt{1 + 4\Lam2 r_+^2} & (2.12c) \cr
}
$$

\noindent where $A_+,A_-,B_+,B_-,C_+,C_-$ are constants determined by the integrated
function with large $r$ limit specified in equation (2.9).  
Once a solution has been found by integration we can find 
$A_+,A_-,B_+,B_-,C_+,C_-$  from
the functions
$f,K,h^{rr}$ and their derivatives.  By varying the initial conditions
we vary the ratio $A_-/A_+$ etc, until these ratios go to zero.  That
would correspond to a regular solution.  We have used a Newton-Raphson
method to investigate this problem and we were not able to find
any regular solution.  Therefore we should  take $f,K,h^{rr}$ to zero.

The next step is to look at the possibility of an instability due to the other  modes.
Taking into account the previous result the equation for $\htr$ becomes
$$
-{(r-r_+)^2\over r^2} {\htr}'' 
-{(2r^2-5rr_+ + 3r_+^2) \over r^3} \htr'
+(\Lambda^2 + {2r^2-6rr_+ + 3r_+^2 \over r^4}) \htr =0
\eqno (2.13)
$$
For $\Lambda r_+ > 1/4$ it is possible to show analytically that 
there is no regular solution to this equation but for smaller
values we have to resort to a numerical analysis similar to 
the one above.  We did not find any regular solution to equation (2.13).
The equation for the mode $h^{rz}$ is similar to
that for $\htr$.  We must therefore take both $h^{rz}$ and $\htr$ to be zero.
With this result we can finally show that $h^{tz}, h^{tt}$ and $h^{zz}$ must
also be zero as their equations are  of the form
$$
-{(r-r_+)^2\over r^2} {h^{tz}}'' 
-g(r) {h^{tz}}'
+\Lambda^2  {h^{tz}}r =0
\eqno (2.14)
$$
where $g(r)$ is an unimportant function of $r$ which can be computed
from
eq (B.11e) of [\paptwo].  These equations
do not possess  regular solutions.  Thus we have shown
that the irregular modes which were found in [\papone] and [\paptwo] do
not have a counterpart for the extreme case.

\noindent{\bf 3. Conclusions.}

In this paper we have shown that the s-mode leading to the instability of
the uncharged and charged p-branes does not extend to the extreme case.  We
have made a detailed search for such a mode and we
were  not able to find one.  This absence of an instability confirms
our expectations based on an extrapolation of the charged p-branes studied in
[\paptwo] and on the entropy argument presented earlier. Black p-branes
therefore behave in a thermodynamically consistent manner.
We should stress that we have not {\it proved} stability, for we have neither
studied higher angular momentum modes, nor exhibited completeness of our basis.
Nonetheless, the fact that the s-mode which gave an unexpected instability
for non- and lesser-charged branes does not give an instability for the
extreme case, and also that adding charge and considering higher angular 
momentum modes has a stabilising effect, 
we regard as compelling evidence for the overall
stability of extremal black p-branes.

Using this result, it is possible to argue that charge
might prevent p-branes from breaking.  When the instability sets in 
in the non-extreme case, the apparent horizon  expands in some regions
and shrinks in others, but the charge per unit length is unchanged, at
least to linear order. Thus it seems possible that these regions will
evolve towards extremality and would therefore be stabilised.  
In this case the endpoint would appear to be a rippled brane, and we are
investigating the possibility of such static solutions.
However, the chargeless
case does not appear to have such a mechanism and we still believe that
these p-branes will break and reveal, at least temporarily, their 
singularity.  

\noindent{\bf Acknowledgments.}

We would like to thank G.W.Gibbons, J.A.Harvey, G.T.Horowitz and P.K.Townsend
for useful  conversations. R.G.~is supported by P.P.A.R.C.
R.L.~thanks Los Alamos National Laboratory for support. 

\parskip=0pt


\newcount\refno
\refno=0
\def\nref#1\par{\advance\refno by1\item{[\the\refno]~}#1}

\def\book#1[[#2]]{{\it#1\/} (#2).}

\def\annph#1 #2 #3.{{\it Ann.\ Phys.\ (N.\thinspace Y.) \bf#1} #2 (#3).}
\def\cmp#1 #2 #3.{{\it Commun.\ Math.\ Phys.\ \bf#1} #2 (#3).}
\def\mpla#1 #2 #3.{{\it Mod.\ Phys.\ Lett.\ \rm A\bf#1} #2 (#3).}
\def\ncim#1 #2 #3.{{\it Nuovo Cim.\ \bf#1\/} #2 (#3).}
\def\npb#1 #2 #3.{{\it Nucl.\ Phys.\ \rm B\bf#1} #2 (#3).}
\def\plb#1 #2 #3.{{\it Phys.\ Lett.\ \bf#1\/}B #2 (#3).}
\def\prd#1 #2 #3.{{\it Phys.\ Rev.\ \rm D\bf#1} #2 (#3).}
\def\prl#1 #2 #3.{{\it Phys.\ Rev.\ Lett.\ \bf#1} #2 (#3).}

\noindent{\bf References.}

 \item{1)} E.Poisson and W. Israel, \prd 41 1796 1990.

\item{2)} G.Gibbons and K.Maeda, \npb 298 741 1988.

D.Garfinkle, G.Horowitz and A.Strominger, \prd 43 3140 1991.

\item{3)} R.Gregory and J.Harvey, EFI 92-49, \prd 47 2411 1993.

J.Horne and G.Horowitz, \npb 399 169 1993.

\item{4)} R.Gregory and R.Laflamme, \prl 70 2837 1993.

\item{5)} R.Gregory and R.Laflamme, \npb 428 399 1994.

\item{6)} G.T.Horowitz and  A.Strominger, \npb 360 197 1991.


\input epsf.tex
\vskip 8truemm
\hskip 0.7truein
\epsfbox{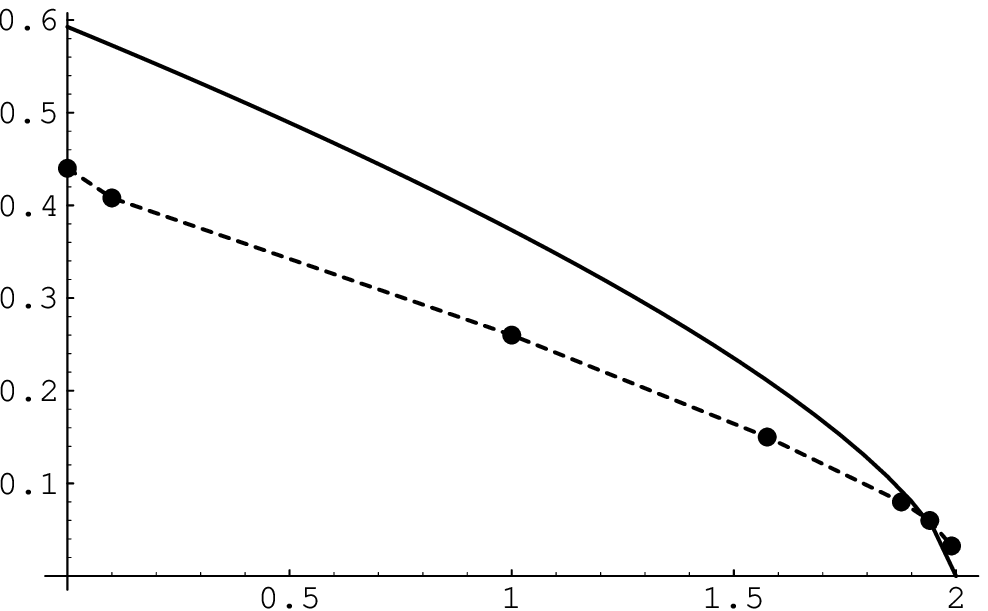}
\vskip -2.truein \hskip 0.2truein $\mu_{{\rm max}}$
\vskip 1.5truein \hskip 3.2truein ${r_-/r_+}$

\noindent {\openup -2\jot
 
\sevenrm Figure 1.  The critical frequency at which the
hyperspherical black hole is entropically preferred to the black string.
The dotted line corresponds to the numerical analysis done in [\paptwo]
and the full one to the analytical argument presented here.  The apparent
crossing of the two curves in the vicinity of  $r_-=r_+$
is probably due to numerical sensitivity. In both cases
the instability does not seem to extend to the extreme case ($r_-=r_+$).

} 
\openup 2\jot  

\bye